\documentclass{PoS}

\title{The hadronic vacuum polarization of the muon from four-flavor lattice QCD}

\ShortTitle{Hadronic vacuum polarization of the muon}

\author{
  \speaker{C.~DeTar} \nolinebreak $^a$
  \thanks{email: \tt detar@physics.utah.edu},
  C.T.H.~Davies$^b$,
  A.X.~El-Khadra$^{cf}$,
  E.~G\'amiz$^d$,
  Steven Gottlieb$^e$,
  D.~Hatton$^b$,
  A.S.~Kronfeld$^f$,
  J.~Laiho$^g$,
  G.P.~Lepage$^h$,
  Yuzhi Liu$^d$,
  P.B.~Mackenzie$^f$,
  C.~McNeile$^i$,
  E.T.~Neil$^j$,
  T.~Primer$^k$,
  J.N.~Simone$^f$,
  D.~Toussaint$^k$,
  R.S.~Van de Water$^f$,
  A.~Vaquero$^a$, and
  Shuhei Yamamoto$^a$
\\
\llap{$^a$} Department of Physics and Astronomy, University of Utah, Salt Lake City, UT 84112 USA\\
\llap{$^b$} SUPA, School of Physics and Astronomy, University of Glasgow, Glasgow, G12 8QQ, UK\\
\llap{$^c$} Department of Physics, University of Illinois, Urbana,  IL 61801, USA\\
\llap{$^d$} CAFPE and Departamento de F\'{\i}sica Te\'orica y del Cosmos, Universidad de Granada, E-18071 Granada, Spain\\
\llap{$^e$} Department of Physics, Indiana University, Bloomington, IN 47405, USA\\
\llap{$^f$} Fermi National Accelerator Laboratory\thanks{Operated by Fermi Research Alliance, LLC, 
  under Contract No.~DE-AC02-07CH11359 with the US DOE.}, Batavia, IL 60510 USA\\
\llap{$^g$} Department of Physics, Syracuse University, Syracuse, NY 13244, USA\\
\llap{$^h$} Laboratory for Elementary-Particle Physics, Cornell University, Ithaca, New York 14853, USA \\
\llap{$^i$} Centre for Mathematical Sciences, University of Plymouth, PL4 8AA, UK \\
\llap{$^j$} Department of Physics, University of Colorado, Boulder, CO 80309, USA\\
\llap{$^k$} Physics Department, University of Arizona, Tucson, AZ 85721, USA\\

\vspace{2mm}
{\large\bf Fermilab Lattice, HPQCD, and MILC Collaborations}
\vspace{2mm}
}

\abstract{We present an update on the ongoing calculations by the
  Fermilab Lattice, HPQCD, and MILC Collaboration of the leading-order
  (in electromagnetism) hadronic vacuum polarization contribution to
  the anomalous magnetic moment of the muon. Our project employs
  ensembles with four flavors of highly improved staggered fermions,
  physical light-quark masses, and four lattice spacings ranging from
  $a \approx 0.06$ to 0.15 fm for most of the results thus far.}

\FullConference{37th International Symposium on Lattice Field Theory - Lattice2019\\
		16-22 June 2019\\
		Wuhan, China}

\begin{document}

\section{Introduction}

The measured anomalous magnetic moment of the muon $a_\mu = (g_\mu -
2)/2$ disagrees with current standard-model predictions by more than
three standard deviations
\cite{Jegerlehner:2017lbd,Davier:2017zfy,Keshavarzi:2018mgv}.  This
discrepancy has generated considerable interest as a possible sign of
new and as-yet-undiscovered fundamental processes.  A concerted effort
is now underway to reduce both theoretical and experimental
uncertainties to clarify the disagreement.  Improved experiments at
Fermilab (E989) and planned at J-PARC (E34) aim to reduce the measured
uncertainty by a factor of about four. To match this on the
theoretical side requires an uncertainty of approximately 0.3\% in the
predicted hadronic contribution, which is the largest source of
theoretical uncertainty.  The most precise theoretical prediction for
this contribution uses dispersion relations together with the measured
cross section for $e^+e^-$ annihilation to hadrons through a single
virtual photon.  The uncertainties in that method become difficult to
control at the level of precision now sought. Because it is an {\it ab
  initio} method, lattice QCD (plus QED) can provide the needed small
uncertainty, but present calculations by several groups are still
quite far from this goal.  Here, we describe progress in the effort of
our collaboration towards reducing the lattice QCD errors to the
target uncertainty. This report updates Ref.~\cite{Davies:2019efs} and
includes new, preliminary results for the quark-line disconnected
contribution.


\section{Formalism}

We use both the time-moment method and time-momentum method
\cite{Chakraborty:2014mwa}, which start from a lattice calculation of
the electromagnetic current-density correlator at zero momentum:
\begin{equation}
  G(t) = \frac{1}{3} \int d{\bf x} \sum_f q_f^2 Z_V^2
  \langle j_i^f({\bf x},t) j_i^f({\bf 0},0)\rangle,
  \label{eq:density-density}
\end{equation}
where the sum is over flavors $f$ with charges $q_f$.  In the
continuum, the reduced vacuum polarization
\begin{equation}
  \hat \Pi(\omega^2) = \frac{4\pi^2}{\omega^2} \int_0^\infty dt
   G(t) \left[\omega^2 t^2 - 4 \sin^2\left(\frac{\omega t}{2}\right)\right]
\end{equation}
determines the leading-order hadronic contribution to anomalous magnetic moment
\begin{equation}
  a_\mu^{\rm HVP,LO} = \left(\frac{\alpha}{\pi}\right)^2 \int_0^\infty dQ^2
  K_E(Q^2) \hat \Pi(Q^2), \label{eq:aHVP}
\end{equation}
where $K_E(Q^2)$ is the Schwinger-term loop integrand as formulated by
Blum \cite{Blum:2002ii}.  The time-moment method expands the reduced vacuum
polarization in powers of $Q^2$
\begin{equation}
  \hat \Pi(Q^2) = \sum_{j=1}^\infty Q^{2j} \Pi_j
\end{equation}
where, on the lattice, the Taylor coefficients are determined by the
discrete time-moments of $G(t)$, {\it i.e.,} $q_f^2 G_{2n} = (-)^n
(2n)! \Pi_{n-1}$, where
\begin{equation}
  G_{2n} \equiv a^4 \sum_{t,{\bf x},f} t^{2n} q_f^2 Z_V^2
      \langle j_i^f({\bf x},t) j_i^f({\bf 0},0)\rangle.
\end{equation}
A Pad\'e approximant is then used to extend the value of $\hat \Pi
(Q^2)$ to small $Q^2$.

Corrections for finite-volume and discretization effects are applied
to the Taylor coefficients $\Pi_{n}$ using a chiral model before
inserting into the integral in Eq.~(\ref{eq:aHVP}).

\begin{figure}
\centering{
  \includegraphics[width=0.43\textwidth]{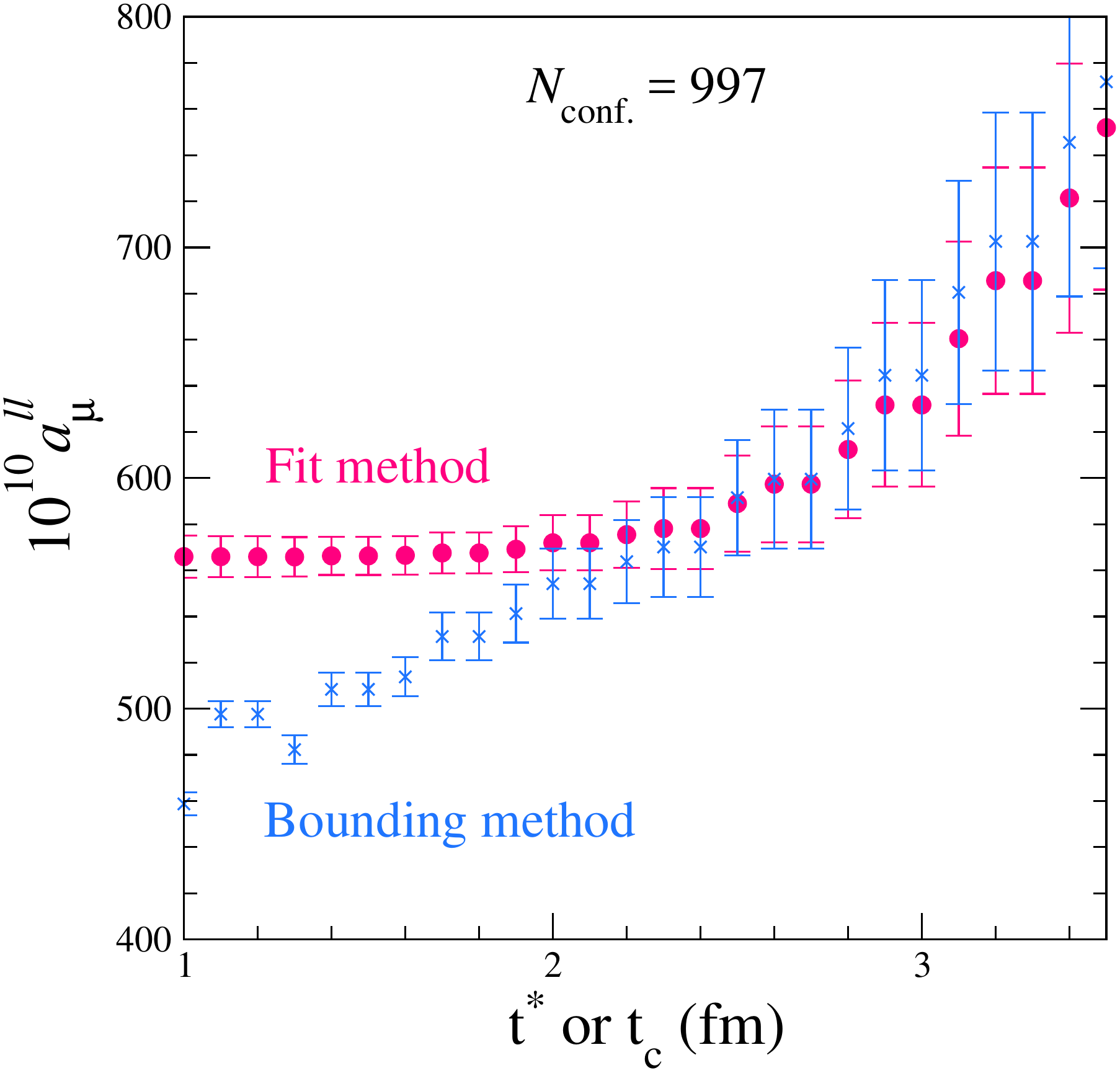}
  \hfill \vspace*{-2mm}
\includegraphics[width=0.46\textwidth]{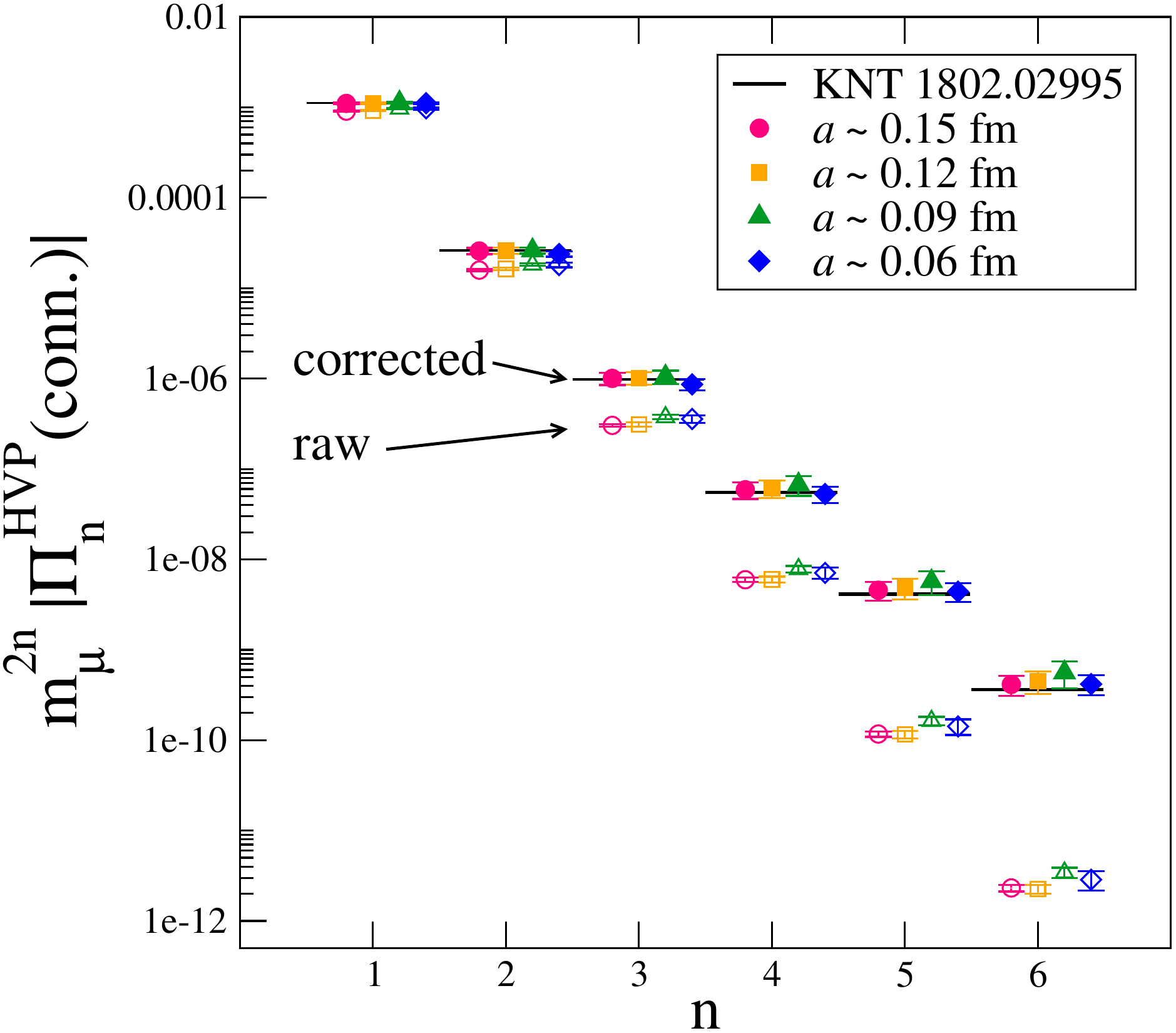}
}
\caption{ Left: Values of $a_\mu^{ll}$ obtained with different methods
  for handling the high $t$ contribution, namely, our ``fit method''
  and the BMW ``bounding method'' \cite{Borsanyi:2016lpl}.  They agree
  well for $t > 2.3$ fm.  Right: Raw (open symbols) and corrected
  (filled symbols) vacuum-polarization Taylor coefficients.  They
  agree well with results from the R-ratio method, shown with black
  lines \cite{Keshavarzi:2018mgv}.
  \label{fig:tcut_and_Pis_vs_R}
}
\end{figure}

\section{Results}

\subsection{Light quark-line-connected contribution}

Our calculations are based on gauge-field ensembles at four lattice
spacings, namely, approximately 0.15, 0.12, 0.09, and 0.06 fm.  These
ensembles were generated in the presence of (2+1+1)-flavors of
highly-improved staggered sea quarks (HISQ) \cite{Follana:2006rc} with
masses close to their physical values \cite{Bazavov:2012xda}.
Measurements are done with the same physical-mass quarks.

The vector-current-density correlator in Eq.~\ref{eq:density-density}
separates into a larger quark-line-connected and a smaller, but more
difficult to calculate, quark-line-disconnected contributions.  We
first discuss the large connected light-quark contribution (with
degenerate up and down quarks).  The largest source of statistical
uncertainty comes from the rapidly growing noise-to-signal ratio in
the current-density correlator at large Euclidean time $t$.  There are
a variety of strategies for treating it.  We fit the
vector-current-density correlator to a set of exponentials and, for
$t$ larger than a cutoff $t^*$, replace the lattice values by an
extrapolation of the fit.  We show in Fig.~\ref{fig:tcut_and_Pis_vs_R}
(left) that for $t^* > 2.3$ fm the result for the contribution to the
anomalous magnetic moment agrees with the BMW bounding method
\cite{Borsanyi:2016lpl}.  One might be concerned that the
extrapolation is risky because our fit model does not include the full
spectrum of states.  We are able to show with a chiral model that the
fit method with only two intermediate states is, in fact, reliable at
the level of accuracy we need \cite{Davies:2019efs}.

\begin{table}
  \caption{Error budget for the quark-line connected light-quark contribution}
\label{tab:errors-ll}  
  \begin{center}
  \begin{tabular}{ll}
   \hline
Lattice spacing uncertainty ($w_0$)           & 0.8\% \\
Monte Carlo statistics                        & 0.7   \\
Continuum extrapolation                       & 0.7   \\
Finite volume and discretization corrections  & 0.6   \\
Current renormalization                       & 0.1   \\
Chiral interpolation                          & 0.1   \\
Strange sea quark mass adjustment             & 0.1   \\
\hline
Total                                         & 1.4\%  \\
\hline
  \end{tabular}
  \end{center}
\end{table}

\begin{figure}
\centering{
\includegraphics[width=0.34\textwidth]{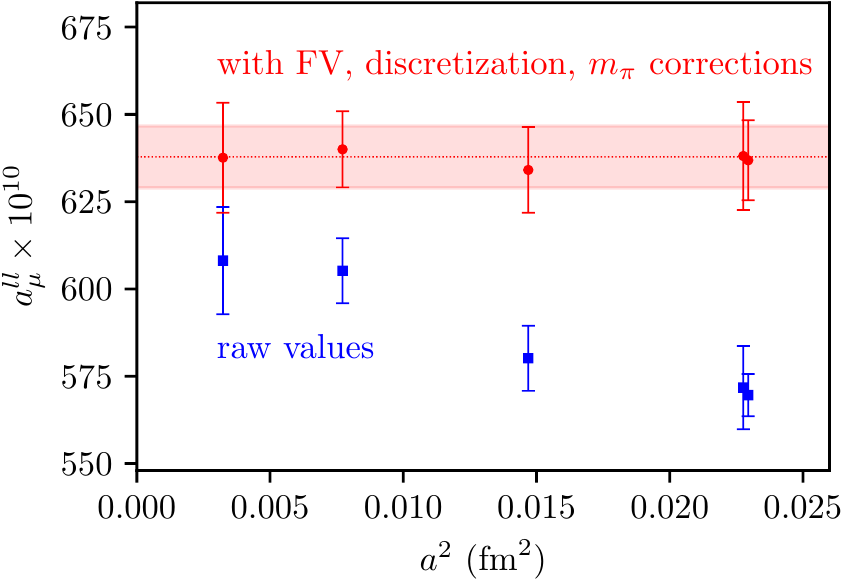} \hfill
\includegraphics[width=0.58\textwidth]{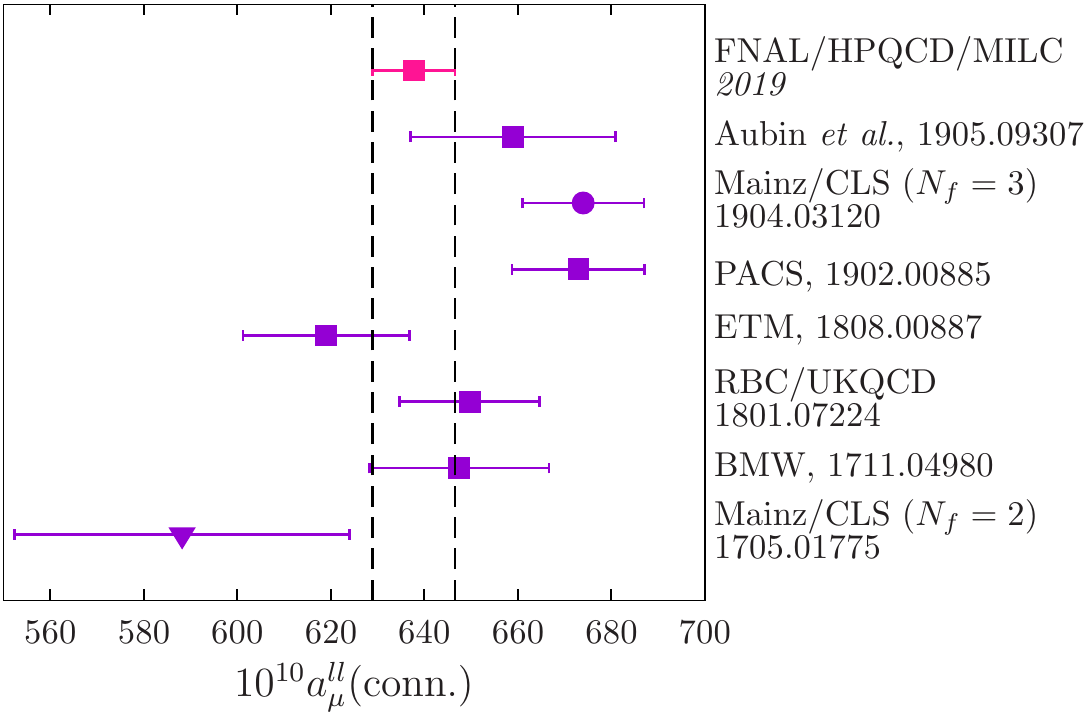}
}
\caption{Left: Continuum extrapolation of our results for the
  light-quark contribution $a_\mu^{ll}$. Shown are values before (open
  blue symbols) and after (filled red symbols) applying corrections
  for finite-volume and discretization effects. Right: Comparison of
  our result for the total light-quark contribution with those of
  other groups
  \cite{Aubin:2019usy,Gerardin:2019rua,Shintani:2019wai,Giusti:2018mdh,Blum:2018mom,Borsanyi:2017zdw,DellaMorte:2017dyu}.
  \label{fig:ll_extrap_compare}
}
\end{figure}

The lattice results require corrections for finite-size and
discretization effects.  A large part of these corrections are
estimated through the use of a chiral model.  The corrected Taylor
coefficients $\Pi_n$ are shown in Fig.~\ref{fig:tcut_and_Pis_vs_R}.
It is noteworthy that our lattice result agrees with the
phenomenological result based on R-ratios \cite{Keshavarzi:2018mgv} at
the $1-2$\% level. The corrected values of $a_\mu^{ll}$ are then
extrapolated to zero lattice spacing, as shown in
Fig.~\ref{fig:ll_extrap_compare}, with the result $a^{ll}_\mu =
637.8(8.8) \times 10^{-10}$, {\it i.e.,} with an error of
1.4\%.\footnote{This result has been updated since the conference.}
The error breakdown is shown in Table~\ref{tab:errors-ll}.  The
dominant contributions to the total error on $a_\mu^{ll}$ are the
lattice spacing uncertainty, statistics, the continuum extrapolation,
and finite volume and discretization corrections. Our result is
compared with recent results of other groups
\cite{Aubin:2019usy,Gerardin:2019rua,Shintani:2019wai,Giusti:2018mdh,Blum:2018mom,Borsanyi:2017zdw,DellaMorte:2017dyu}
in Fig.~\ref{fig:ll_extrap_compare}.

\begin{table}
  \caption{Adjustments to obtain the total up/down quark contribution
    to $a_\mu^{ud}$.  All values are in units of $10^{-10}$.}
\label{tab:get_ud}
\begin{center}
\begin{tabular}{lr}
   \hline
$M(\pi^0) - M(\pi^+)$              &  $-4.3$  \\
$\pi - \pi$ disconnected           &  $-7.9$  \\
\hline
Total                              & $-12(3)$ \\
\hline
disconnected                       &  $-5(5)$ \\
Strong isospin breaking            & $10(10)$ \\
Electromagnetism                   &   $0(5)$ \\
\hline
Total adjustment                   & $-7(13)$ \\
   \hline
  \end{tabular}
\end{center}
\end{table}

\subsection{Total leading-order light-quark hadron vacuum polarization}

To obtain the total leading-order up/down quark contribution to
$a_\mu$ we need to include strong isospin-breaking and electromagnetic
effects, and we need the quark-line-disconnected contribution.  A
large part of the strong isospin-breaking comes from the splitting of
the pion multiplet in the two-pion intermediate state.  This
contribution can be calculated from the same chiral model used in the
finite-volume and lattice-discretization correction.  The same
procedure can also be used to estimate the quark-line-disconnected
contribution. (Our more recent, explicit calculation, discussed below,
was not used in estimating this contribution.)  The residual
(non-pion-pion) contributions are then estimated conservatively, as
shown in Table~\ref{tab:get_ud}.  In the case of strong isospin
breaking, explicit lattice calculations help provide the estimates
shown  \cite{Blum:2018mom,Chakraborty:2017tqp,Giusti:2019xct}.  The
largest uncertainty here still comes from strong isospin
breaking. Altogether, then, the total up-down quark contribution is,
then, $a_\mu^{ud} = 630.8(8.8)(13) \times 10^{-10}$, where the first
error comes from the light-quark value above and the second from the
corrections in Table~\ref{tab:get_ud}.

\begin{figure}
\centering{
  \includegraphics[width=0.60\textwidth]{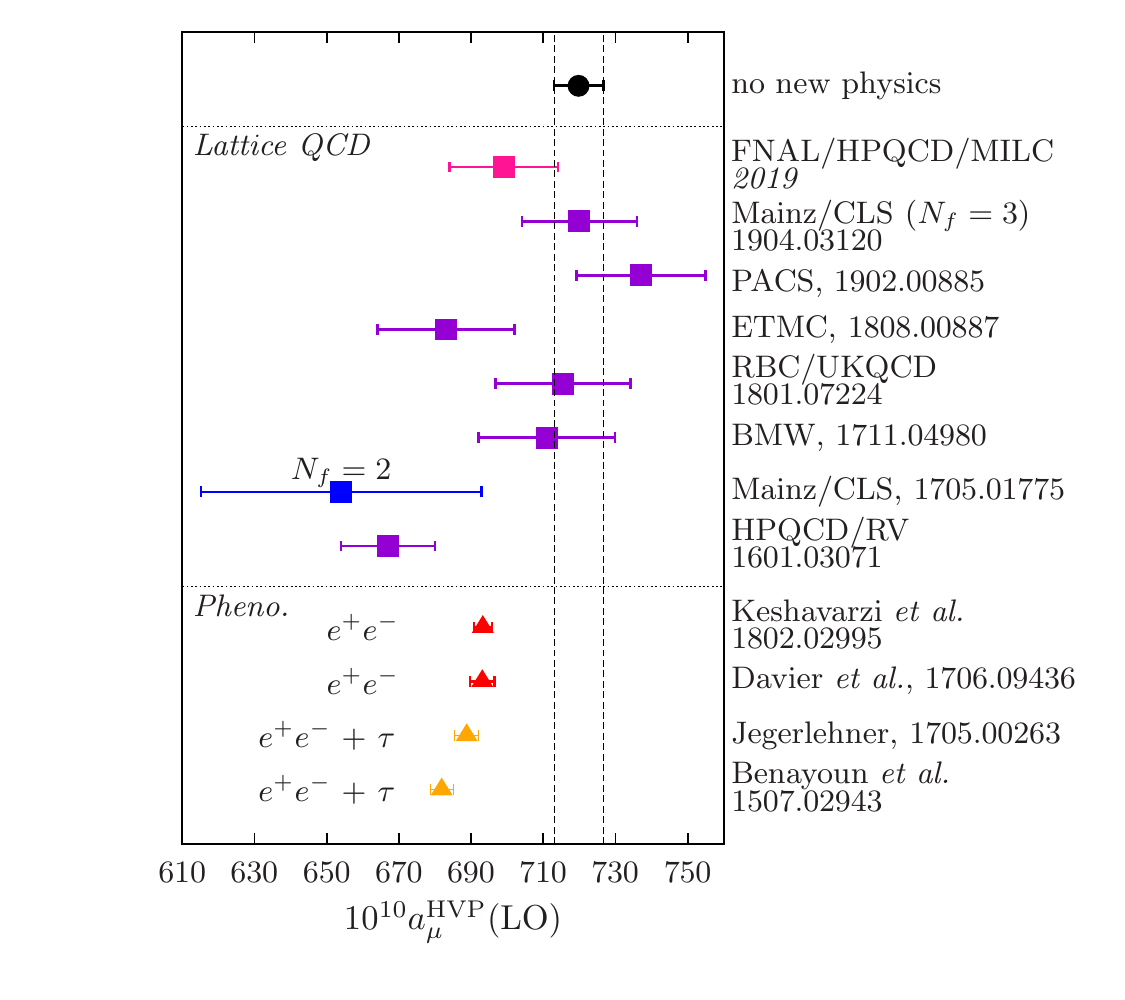}
}
\vspace*{-4mm}
\caption{Comparison of results for the total leading-order hadronic
  vacuum polarization contribution to $a_\mu^{\rm LO,HVP}$
  \cite{Jegerlehner:2017lbd,Davier:2017zfy,Keshavarzi:2018mgv,Gerardin:2019rua,Shintani:2019wai,Giusti:2018mdh,Blum:2018mom,Borsanyi:2017zdw,DellaMorte:2017dyu,Chakraborty:2016mwy,Benayoun:2015gxa}.
  \label{fig:compare_tot}
}
\end{figure}

Finally, including further contributions from the strange, charm, and
bottom quarks
\cite{Chakraborty:2014mwa,Donald:2012ga,Colquhoun:2014ica} gives
$a^\mu_{\rm HVP-LO} = 699(15)_{u,d}(1)_{s,c,b} \times 10^{-10}$, where
the errors are from the light-quark and heavier-quark contributions,
respectively.  The largest contribution to the uncertainty comes from
isospin-breaking and electromagnetic corrections.  This value is
compared with those of others in Fig.~\ref{fig:compare_tot}.

\begin{figure}
\centering{
  \includegraphics[width=0.60\textwidth]{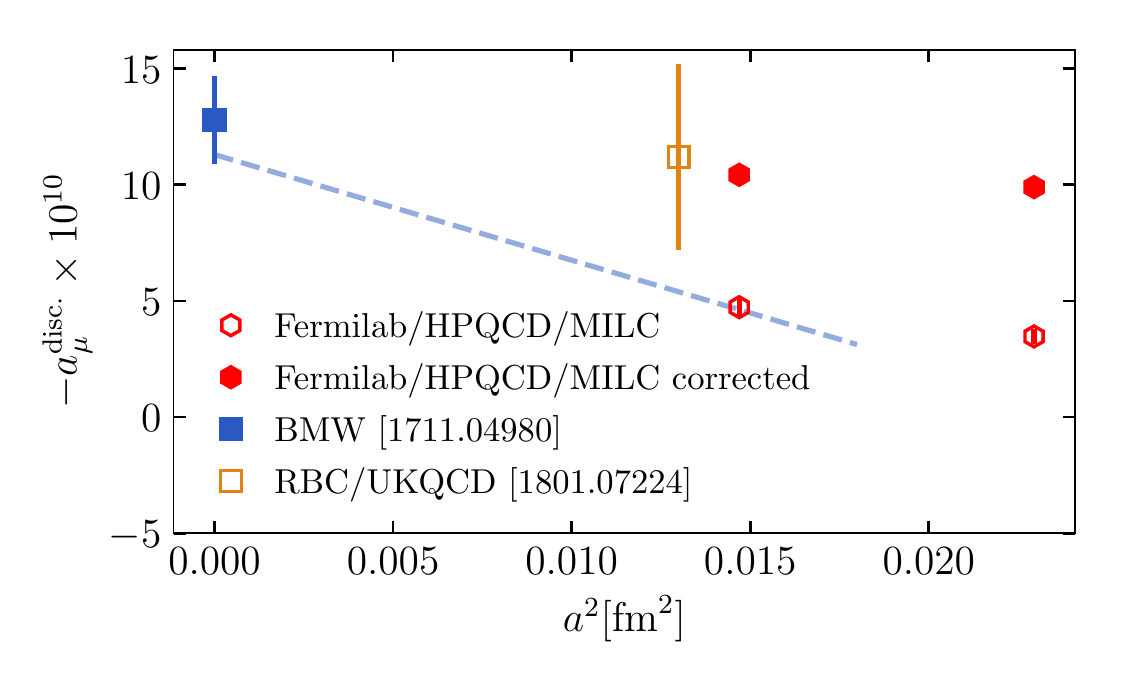}
}
\vspace*{-4mm}
\caption{ Our {\em preliminary} results for the quark-line disconnected
  contribution to $a_\mu^{\rm LO,HVP}$, compared with those of the BMW
  \cite{Borsanyi:2017zdw} and RBC/UKQCD \cite{Blum:2018mom} 
  collaborations. Our results are shown before and after applying a
  correction for finite-volume and discretization effects.
  \label{fig:disconn-extrap}
}
\end{figure}

\section{Quark-line-disconnected HVP}

We conclude by presenting some preliminary results from our explicit
lattice calculation of the quark-line-disconnected contribution, which
will eventually replace the model estimates described above. So far we
have results for simulations at only two lattice spacings 0.15 and
0.12 fm, updated from work described in
Ref.~\cite{Yamamoto:2018cqm}. The resulting contributions are
summarized in Fig.~\ref{fig:disconn-extrap} and compared with results
from the BMW \cite{Borsanyi:2017zdw} and RBC/UKQCD Collaborations
\cite{Blum:2018mom}.  The results are corrected for finite-volume and
discretization effects, which removes most of the lattice-spacing
dependence.  We see that the finite-volume plus discretization-error
correction is significant.  Although the extrapolated value is
statistically consistent with the model estimates described above, a
calculation at a finer lattice spacing is evidently needed in order to
meet our goals for a reduction in the overall error.

\section{Outlook}

We continue in our efforts to reduce the lattice uncertainty.  Besides
increasing statistics and working at finer lattice spacing, we are
also working to reduce the uncertainties in the lattice scale, we are
planning calculations of electromagnetic corrections, and we are
investigating lattice calculations of contributions from two-hadron
intermediate states.

\section*{Acknowledgments}

This work is supported by grants from the US Department of Energy, the
US National Science Foundation, the United Kingdom STFC, MINECO
(Spain), the Junta de Andaluc\'ia (Spain), the German Excellence
Initiative and the European Union Seventh Framework Program, the
European Union's Marie Curie COFUND program, and the Blue Waters PAID
program.  Computations were carried out on USQCD LQCD clusters and at
the US DOE NERSC and ALCF centers, the U.K. STFC DiRAC HPC Facility,
the US NSF TACC XSEDE center, and computers at the University of
Colorado and Indiana University.

\end{document}